\documentclass[preprint,prb,showpacs,preprintnumbers,amsmath,amssymb]{revtex4}
\usepackage{latexsym}
\usepackage{amssymb}
\usepackage{amsmath}
\usepackage{graphicx}
\usepackage{natbib}

\usepackage{color}
\usepackage{bm}

\newcommand{\divv}{\mathop{\rm div}\nolimits}
\newcommand{\grad}{\mathop{\rm grad}\nolimits}

\begin{document}
\title{Magnetic field effect on polarization and dispersion of exciton-polaritons in planar microcavities }
\author{D. D. Solnyshkov}
\affiliation{LASMEA, UMR 6602 CNRS, Universit\'{e} Blaise-Pascal, 24, av des Landais, 63177, Aubi\`{e}re, France.}
\author{M.M. Glazov}
\affiliation{A.F. Ioffe Physico-Technical Institute, 26, Politechnicheskaya, 194021, St-Petersburg, Russia}
\author{I.A. Shelykh}
\affiliation{Science Dept., University of Iceland, Dunhaga-3, IS-107, Reykjavik, Iceland}
\affiliation{ICCMP, Universidade de Brasilia, 70904-970, Rrasilia-DF, Brazil}
\author{A.V. Kavokin}
\affiliation{Department of Physics and Astronomy, University of Southampton, SO17 1BJ Southampton, United Kingdom}
\affiliation{Marie Curie Chair of Excellence POLAROMA, University of ROme II TOr Vergata, 1, via della Ricerca Scientifica, 00133, Rome, Italy}
\author{E.L. Ivchenko}
\affiliation{A.F. Ioffe Physico-Technical Institute, 26, Politechnicheskaya, 194021, St-Petersburg, Russia}
\author{G.~Malpuech}
\affiliation{LASMEA, UMR 6602 CNRS, Universit\'{e} Blaise-Pascal, 24, av des Landais, 63177, Aubi\`{e}re, France.}

\date{\today}

\begin{abstract}
The non-local dielectric response theory is extended to describe
oblique reflection of light from quantum wells subjected to the
magnetic field. This allows us to calculate the dispersion and
polarization of the exciton-polariton modes in semiconductor
microcavities in the presence of a magnetic field normal to the
plane of the structure. We show that due to the interplay between
the exciton Zeeman splitting and TE-TM splitting of the photon
modes, four polariton dispersion branches are formed with a
polarization gradually changing from circular in the exciton-like
part to linear in the photon-like part of each branch. Faraday rotation in quantum microcavities is shown to be strongly enhanced as compared with the rotation in quantum wells.
\end{abstract}

\maketitle

\section{Introduction}

Cavity polaritons are half-light half-matter quasiparticles
resulting from the strong coupling of the photon mode of a
microcavity with an exciton resonance of the embedded
semiconductor structure~\cite{See:2003}. 

Being mixed exciton-photon quasiparticles, the cavity polaritons
have integer spins and can reveal bosonic
properties~\cite{Richard:2005} responsible for a number of
interesting effects both predicted and observed, namely,
stimulated scattering~\cite{Savvidis:2000}, polariton lasing~\cite{Imamoglu:1996}, Bose-Einstein
condensation~\cite{Kasprzak:2006,Christop},
superfluidity~\cite{Shelykh:2006,Carusotto:2004} etc. Many
important properties of the polaritonic systems are connected with
their spin degree of freedom~\cite{Shelykh:2005}. It was shown
that an external magnetic field strongly affects non-linear
dynamics of the polaritons and optical properties of polariton
condensates. Recently, the magnetic field induced suppression of
polariton superfluidity and the so-called ``spin Meissner effect''
have been predicted~\cite{Rubo:2006}.

For understanding of the rich variety of non-linear phenomena
induced by magnetic field in microcavities a thorough knowledge of
the linear properties of magneto-polaritons (i.e. polaritons subject to a magnetic field) is needed. Though the
problem on the surface seems simple it becomes really complex if
the polarization of light is included into consideration. In
particular, to the best of our knowledge, till now there was no
accurate analysis of the dependence of eigenenergies of
exciton-polariton modes and their polarization on the magnetic
field. The present paper is aimed at filling the apparent gap in
the theory of cavity polaritons. We expand the non-local
dielectric response theory to describe oblique reflection and
transmission of light through quantum wells (QWs) in the presence
of the magnetic field normal to the QW plane. We present also the
transfer-matrix calculation of the eigenenergies and polarization
of exciton-polaritons accounting for the TE-TM splitting of cavity
modes. Finally, Faraday rotation of the polarization plane of light passing through the microcavity is analyzed.

The polarization of cavity polaritons depends on many factors. In
the absence of the external magnetic field the exciton-polariton
states having a non-zero in-plane wave vector ($k\ne 0)$ are
linearly polarized due to the long range electron-hole exchange
interaction~\cite{Maialle:1993} and TE-TM splitting of the
photonic mode of the cavity~\cite{Panzarini:1999}. Application of
a magnetic field normal to the QW plane (Faraday geometry) results
in the Zeeman splitting of the exciton doublet and thus changes
the polarization of the polariton eigenstates from linear to
elliptical or circular. As the bare photonic modes are unaffected
by the magnetic field, the degree of circular polarization of
exciton-polariton states is strongly dependent on the relative
weights of photonic and excitonic components within the given
polariton state. In strong magnetic fields these weights are
field-dependent due to the shrinking of the wavefunction of the
relative motion of the electron and hole in the real space
resulting in an increase of the exciton-photon coupling strength.
The interplay between these phenomena has not been addressed in
its complexity to the best of our knowledge.

Experimentally, the magnetic field effect on the spectrum of
exciton-polaritons in microcavities has been studied in 1990s by
several groups~\cite{Tignon:1997,Berger:1996,Armitage:1997}. The
magnetic field induced weak-strong coupling transition has been
observed, and the enhancement of the Rabi-splitting of the
exciton-polariton modes with the magnetic field increase has been
measured by means of {\it cw} and time-resolved optical
spectroscopies. The shape and polarization of the polariton
dispersion curves have not been studied, however. Later on, mixing
of the bright-polariton and dark exciton states due to the
magnetic field applied in the cavity plane has been studied using
the polarized photoluminescence~\cite{Renucci:2003} and
time-resolved Kerr rotation~\cite{Brunetti:2006,Brunetti:2007}
techniques. Theoretically, the dispersion of exciton-polaritons in
microcavities in the absence of the magnetic field has been
described in 1990s~\cite{Savona:1995,Kavokin:1995}, the magnetic
field effect on the exciton-photon coupling strength has been
calculated in~\cite{Armitage:1997}, and the exciton-polariton
longitudinal-transverse splitting has been studied in detail in~\cite{Panzarini:1999}.

The present paper is organized as follows. In the second section
we calculate the reflectivity of a QW structure in the presence of
the external magnetic field applied perpendicular to the well
plane. In the Section III we calculate the dispersion of
magneto-exciton-polaritons in a microcavity and analyze their
polarization. In Section IV Kerr and Faraday effects in microcavities are discussed.

\section{Reflection of light from a QW in the magnetic field}
For a heavy-hole exciton usually being the lowest energy exciton
state in a QW, the allowed spin projection $J_z $ to the structure
growth axis is either $\pm 2$ or $\pm 1$ depending on the mutual
orientation of the electron and hole spins. The states with $J_z
=\pm 2$ are decoupled from the cavity modes while the coupling of
the $\pm 1$ states with the right or left circularly polarized
cavity photons gives rise to the polariton doublets. In the
following we will bound ourselves to consider only the bright
states $J_z = \pm 1$.

Let us consider the reflection of light incident at oblique angle
on a QW which is sandwiched between the semi-infinite barriers and subject to the magnetic field $\bm{B}$ normal to
its plane. We shall take into account the Zeeman splitting of the
exciton resonance and neglect for a while the magnetic field
effect on the orbital motion of electron and hole in the exciton.
Then the resonance frequencies of the $+1$ and $-1$ excitons can
be presented respectively by $\omega _\pm = \omega_0(k) \pm g \mu_B B/2 \hbar$,
where $\mu_B$ is the Bohr magneton, $g$ is the exciton
$g$-factor, $\omega_0(k)$ is the exciton resonance frequency given
by
\begin{equation}\label{eq:exc} \hbar \omega_0(k) = E_g^{QW} -
E_B + \frac{\hbar^2 k^2}{2M}\:,
\end{equation}
$E_g^{QW}$ is the effective band gap of the QW calculated with
allowance for size-quantization energy, $E_B$ is the exciton
binding energy, $M$ is the exciton translational mass and $k$ is
its in-plane wave vector. 
%The exciton frequency $\omega_0(k)$ can
%depend on the magnetic field due to diamagnetic orbital effect of
%the magnetic field discussed in Section IV.

%\textcolor{blue}{Misha: We removed all this discussions of orbital effects because they are very important in the case of microcavities. EL, can I ask you to write a few words about diamagnetic shift and modification of the oscillator strength in any appropriate place?}

The electric field of the electromagnetic wave is presented in the
standard form $\bm{E}(\mathbf r) {\rm e}^{- {\rm i} \omega t} +
\bm{E}^*(\mathbf r) {\rm e}^{{\rm i} \omega t}$, where the light
frequency $\omega$ lies in the vicinity of the QW exciton
resonance. The wave equation for the vector $\bm{E}(\mathbf r)$
can be written as
\begin{equation}
\label{eq1}
\nabla^2{\bm E} + q^2 {\bm E} = - 4 \pi \left( {\frac{\omega }{c}}
\right)^2\left( {{\bm P} + \frac{1}{q^2}\grad \divv{\bm P}}
\right)\:,
\end{equation}
where $q = {\omega \sqrt \varepsilon } \mathord{\left/ {\vphantom
{{\omega \sqrt \varepsilon } c}} \right.
\kern-\nulldelimiterspace} c$, $\varepsilon $ is the background
dielectric constant of the material surrounding the QW, $\bm{P}$
is the exciton polarization induced by the electromagnetic wave.
In our further consideration we denote the structure growth axis
as $z$, and suppose that the light propagates in the $xz$ plane.
The electric field of TE-polarized light is parallel to the
$y$-axis while the electric field of TM-polarized light lies in
the $xz$ plane. Let $\theta$ be the incidence angle, i.e. the angle between $z$ axis and light wavevector. Representing the solution in the form ${\bm
E}(\mathbf r) = {\bm E}(z)e^{{\rm i} q_x x}$, taking into account
that $q_x = k_x \equiv k$ and using one-dimensional Green's
function of the wave equation $G(z,z')= \left( {{\rm i} /2q_z } \right)\exp{( {\rm i}
q_z \left| {z-z'} \right|)}$, one can reduce Eq. (\ref{eq1}) to
the following integral equation
\begin{equation}
\label{eq2}
{\bm E}= {\bm E}_0 e^{{\rm i} \left( {q_z z + k_x x} \right)} + \frac{2 \pi
{\rm i}}{q_z }\left( {\frac{\omega }{c}} \right)^2\int\limits_{-\infty }^{+\infty
} {\left( {{\bm P} + \frac{1}{q^2}\grad \divv{\bm P}} \right)}
e^{{\rm i}q_z \left| {z-z'} \right|}dz'\:.
\end{equation}
This equation should be completed by the material relation linking
the electric field and polarization. The latter can be written
assuming the nonlocal dielectric response of the QW in the exciton
resonance frequency region~\cite{Andreani:1991,Ivchenko:1992}. In order to do so, we remark that the 
amplitudes of the right and left circularly polarized components,
$E_{\pm}$ or $P_{\pm}$, are related to those of linearly polarized
components by
\begin{equation}
\label{eq3}
E_{\pm} =E_x \mp {\rm i} E_y, \quad P_{\pm} = P_x \mp {\rm i} P_y\:.
\end{equation}
Note that we consider only the polarization induced by heavy-hole
exciton and do not take into account the $z$-component of the
excitonic polarization which may be induced due to a light-hole
exciton~\cite{Tassone:1992}, so that $P_z \equiv 0$.

In the basis of circular polarized components the material
equation has the form
\begin{equation}
\label{eq5}
4 \pi P_\pm (z) = \frac{Q \Phi (z)}{\omega _\pm - \omega - {\rm i} \gamma
}\int\limits_{- \infty }^{+\infty } {\Phi (z')} E_\pm (z')dz' = \frac{Q \Phi (z)
\Lambda_\pm}{\omega_\pm - \omega - {\rm i} \gamma} \quad\: ,
\end{equation}
Here $\Phi(z)$ is the exciton wave-function taken with equal
electron and hole coordinates,
\begin{equation}
\label{eq7}
\Lambda_j = \int\limits_{- \infty}^{+ \infty} {\Phi (z')} E_j (z')dz'\:,
\end{equation}
$\gamma$ is the homogeneous broadening of the exciton resonance
caused by the acoustic phonon scattering and the QW imperfections,
$Q=\varepsilon \omega_{LT} \pi a_B^3 $ with $\omega _{LT} $ and
$a_B $ being the longitudinal-transverse splitting and Bohr radius
of bulk excitons, respectively.

Equations \eqref{eq5} represent the generalization of the
expression for the dielectric polarization within the framework of
the nonlocal model of the dielectric response for QWs with a
spin-degenerate exciton resonance. The condition $P_z = 0$ allows
one to decouple the equation for the $z$-component of the
electromagnetic field from those for the $x$- and $y$-components
which read
\begin{equation}\label{eq6a}
E_x(z) = E_{0x} {\rm e}^{{\rm i} q_z z} + {\rm i} \frac{\pi Q q_z}{q^2} \left( {\frac{\omega
}{c}} \right)^2 \left( \Sigma_+ + \Sigma_- \right) {\int\limits_{-\infty }^{+\infty } {\Phi (z')} {\rm e}^{{\rm i} q_z
\left| {z-z'} \right|}dz'} \:,
\end{equation}
\[
E_y(z) = E_{0y} {\rm e}^{{\rm i} q_z z} - \frac{\pi Q}{q_z} \left( {\frac{\omega
}{c}} \right)^2 \left( \Sigma_- - \Sigma_+ \right) {\int\limits_{-\infty }^{+\infty } {\Phi (z')} {\rm e}^{{\rm i} q_z
\left| {z-z'} \right|}dz'} \:,
\]
where
\begin{equation}
 \label{sigmas}
\Sigma_+ = \frac{\Lambda_x - {\rm i} \Lambda_y }{\omega _+ -\omega - {\rm i} \gamma},
\quad \Sigma_- = \frac{\Lambda _x + {\rm i} \Lambda _y }{\omega _- -\omega - {\rm i} \gamma }\:.
\end{equation}
Multiplying both parts of Eq. \eqref{eq5} by $\Phi (z)$ and
integrating over $z$ from $-\infty $ to $+\infty $ one can obtain
a closed set of equations for $\Lambda _{x,y}$
\begin{equation}
\label{eq9}
\Lambda_x = \Lambda_{0x} + \frac{q_z}{q}\frac{ {\rm i} \Gamma _0 - \Delta \omega_p
}{2}\left( \Sigma_+ + \Sigma_- \right)\:,
\end{equation}
\[
\Lambda_y = \Lambda_{0y} + {\rm i} \frac{q}{q_z} \frac{- \Delta \omega_p + {\rm i} \Gamma
_0 }{2} \left( \Sigma_- - \Sigma_+ \right)\:.
\]
Here
\begin{equation}
\label{eq11}
\Delta \omega_p =\frac{2 \pi Q}{q}\left( {\frac{\omega }{c}} \right)^2
{\int\limits_{-\infty }^{+\infty } {\int\limits_{-\infty }^{+\infty } {\Phi
(z')} \Phi (z)} \sin{( q_z \left| {z-z'} \right|)}dz'dz}
\end{equation}
is a renormalization shift of the exciton resonance frequency due
to the light-matter coupling effect at normal incidence in the
absence of the external magnetic field,
\begin{equation}
\label{eq12}
\Gamma _0 = \frac{2 \pi Q}{q} \left( {\frac{\omega }{c}} \right)^2
{\int\limits_{-\infty }^{+\infty } {\int\limits_{-\infty }^{+\infty } {\Phi
(z')} \Phi (z)} \cos{( q_z \left| {z-z'} \right|)}dz'dz}
\end{equation}
is the exciton radiative broadening in the same conditions, and
\begin{equation}
\label{eq9_v}
\Lambda _{0j} = E_{0j} \int\limits_{- \infty}^{+ \infty} {\Phi (z')} {\rm e}^{{\rm i} q_z
z'}dz' \quad (j=x,y)\:.
\end{equation}

In general, both the exciton resonance frequency and oscillator strength depend on the magnetic field due to the orbital motion of electron and hole. The magnetic field presses an electron and a hole to each other thus increases its binding energy and oscillator strength~\cite{Kavokin:1993,ivchbook}. These effects are neglected in the following calculation but they can be readily taken into account by renormalizing the values of $\omega_0$ and $\Gamma_0$.

\begin{figure}[htbp]
\centerline{\includegraphics[width=\linewidth]{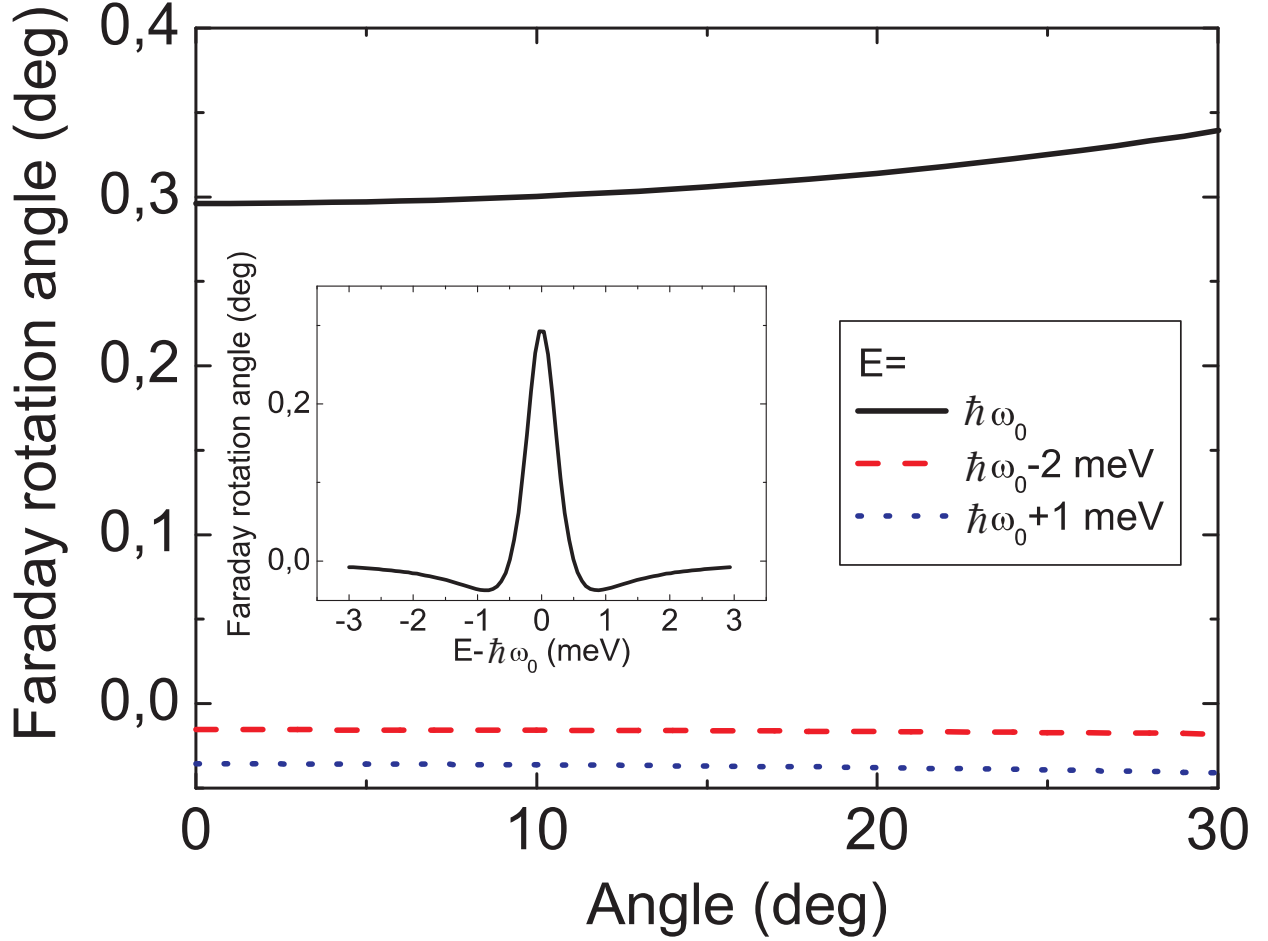}}
\caption{Faraday rotation angle by a single QW versus the angle of incidence for different
energies close of the exciton resonance. $\hbar \omega_0 = 1.4$~eV, $\hbar \Gamma_0 = 0.026$~meV, $\hbar \gamma=0.5$~meV, Zeeman splitting is 0.1 meV. Inset shows Faraday rotation angle calculated as a function of energy for the same quantum well at normal light incidence.}\label{fig:faraday}
\end{figure}

The solution of linear equations (\ref{eq9}) can be presented in
the form
\begin{equation}
\label{eq13}
\Lambda _x =\alpha _x \Lambda _{0x} + \alpha _y \Lambda _{0y}\:,
\end{equation}
\[\Lambda_y =\beta _x \Lambda _{0x} + \beta _y \Lambda _{0y}\:,
\]
where
\begin{eqnarray}
\label{eq14}
&&\alpha _x =\left( {1+\frac{q_z }{2q}\frac{\Delta \omega_p - {\rm i} \Gamma _0
}{\Omega^+ }} \right) \frac{1}{\mathcal D}\:, \quad \alpha _y = \frac{{\rm i} q_z }{2q}\frac{\Delta \omega_p - {\rm i}
\Gamma_0 }{\mathcal D \Omega^- } \:, \\
&&\beta_x =-\frac{iq}{2 q_z }\frac{\Delta \omega_p - {\rm i} \Gamma _0 }{\mathcal D \Omega^-} \:,
\quad \beta_y =\left( {1 + \frac{q}{2 q_z }\frac{\Delta \omega_p - {\rm i} \Gamma_0
}{\Omega^+ }} \right) \frac{1}{\mathcal D}\:, \nonumber \\
&&\mathcal D = \left( {1 + \frac{q}{2q_z }\frac{\Delta \omega_p - {\rm i} \Gamma _0 }{\Omega^+ }}
\right)\left( {1 + \frac{q_z }{2q}\frac{\Delta \omega_p - {\rm i} \Gamma _0 }{\Omega^- }} \right) -
\left( {\frac{\Delta \omega_p - {\rm i} \Gamma _0 }{2\Omega^- }}
\right)^2\:,\nonumber
\end{eqnarray}
and $(\Omega^{\pm})^{-1} = (\omega _+ -\omega - {\rm i}
\gamma)^{-1} \pm (\omega_- - \omega - {\rm i} \gamma)^{-1}$. In
order to find the reflection and transmission coefficients of the
QW we should find the asymptotic values of the electric field
far beyond the QW. Coming back to Eq. (\ref{eq6a}) and putting $z
\to \pm \infty $ one easily obtains
\begin{eqnarray}
\label{eq20}
&&E_x (z \to \pm \infty)=E_{0x} {\rm e}^{{\rm i}q_z z} + {\rm i} {\rm e}^{\pm {\rm i} q_z z} \frac{\pi Q q_z}{q^2}\left(
{\frac{\omega }{c}} \right)^2 \left( \Sigma_+ + \Sigma_- \right) {\int\limits_{-\infty }^{+\infty } {\Phi
(z')} e^{\mp {\rm i}q_z z'}dz'} \:, \\
&&E_y (z \to \pm \infty) = E_{0y} {\rm e}^{{\rm i} q_z z} - \hspace{2.5 mm}{\rm e}^{\pm {\rm i} q_z z} \frac{\pi Q}{q_z}\left(
{\frac{\omega }{c}} \right)^2 \left( \Sigma_- - \Sigma_+\right) {\int\limits_{-\infty }^{+\infty } {\Phi
(z')} e^{\mp {\rm i} q_z z'}dz'} \:. \nonumber
\end{eqnarray}
The transmission and reflection coefficients are
polarization-dependent and can be represented in the form of
$2\times 2$ matrices $\hat r$ and $\hat t$ defined by
\begin{equation}
\label{eq24}
{\bm E}(z \to -\infty) -  {\bm E}_0 e^{{\rm i} q_z z}
= \hat{r}  {\bm E}_0 e^{{\rm i} q_z z}\:,\:
{\bm E} (z \to + \infty)  = \hat{t}
{\bm E}_0 e^{{\rm i} q_z z}\:.
\end{equation}
Using Eqs. (\ref{eq13})--(\ref{eq24}) one obtains after simple algebra the reflection
matrix elements,
\begin{eqnarray}
\label{eq26}
&&r_{xx} =\frac{{\rm i} \Gamma _0 }{2}\left( {\frac{1}{\Omega^+ } + \frac{q}{q_z
}\frac{\Delta \omega_p - {\rm i} \Gamma _0 }{\left( {\omega_+ - \omega - {\rm i} \Gamma}
\right) \left( {\omega_- - \omega - {\rm i} \Gamma} \right)}} \right)\:,\\
&&r_{yy} =\frac{{\rm i} \Gamma _0 }{2}\left( {\frac{1}{\Omega^+ } + \frac{q_z
}{q} \frac{\Delta \omega_p - {\rm i} \Gamma _0 }{\left( {\omega_+ - \omega - {\rm i} \Gamma
} \right)\left( {\omega_- - \omega - {\rm i}\Gamma } \right)}} \right)\:, \nonumber \\
&&r_{xy} = - r_{yx} = \frac{\Gamma _0 }{2\Omega^- \mathcal D}\:. \nonumber
\end{eqnarray}
The transmission and reflection coefficients are interrelated by
\begin{equation}
\label{eq28}
t_{ij} =\delta_{ij} + r_{ij}\:,
\end{equation}
where $\delta_{ij}$ is the Kronecker delta. Note that, due to the
Zeeman splitting of the exciton resonance into a circularly
polarized doublet, the coupling of light with an exciton leads to
the mixing of TE and TM polarizations both in reflection and
transmission. This results in the resonant Faraday and Kerr
rotation and dichroism~\cite{Brunetti:2006}.

Figure \ref{fig:faraday} shows the angle of rotation of the
polarization plane of transmitted light with respect to the polarization plane of incident light
(which is assumed to be TE-polarized) versus the angle of incidence for different
energies close of the exciton resonance for the  magnetic field of
2 T (the induced Zeeman splitting is 0.1 meV). We have considered
a 10 nm GaAs QW with $\Gamma_0=0.026$~meV.  The homogeneous
broadening is 0.5 meV. The effect includes both linear-circular
dichroism and Faraday rotation. The angle of Kerr rotation
can be calculated by using the following equation, see, e.g.,
\cite{LandauLif},
\begin{equation}
\theta_K = \frac12 \arctan{\frac{2 {\rm Re}\{r_{yy}^* \tilde r_{xy}\}}{|r_{yy}|^2 - |\tilde  r_{xy}|^2}}
\:.
\end{equation}
Here $\tilde r_{xy} = r_{xy}/\cos{\theta}$.
The Faraday rotation angle $\theta_F$ is obtained by the replacement $r_{ij}
\to t_{ij}$. For small rotations a simplified equation $\theta_K
\approx {\rm Re} \{ r_{xy} / (r_{yy}\cos{\theta}) \}$ can be applied. One can
check that the effect is strongest at the resonance and decreases
with the increasing angle of incidence.

\section{Dispersion of exciton-polaritons in microcavities subjected to the magnetic field}

\begin{figure}[htbp]
\centerline{\includegraphics[angle=270,width=\linewidth]{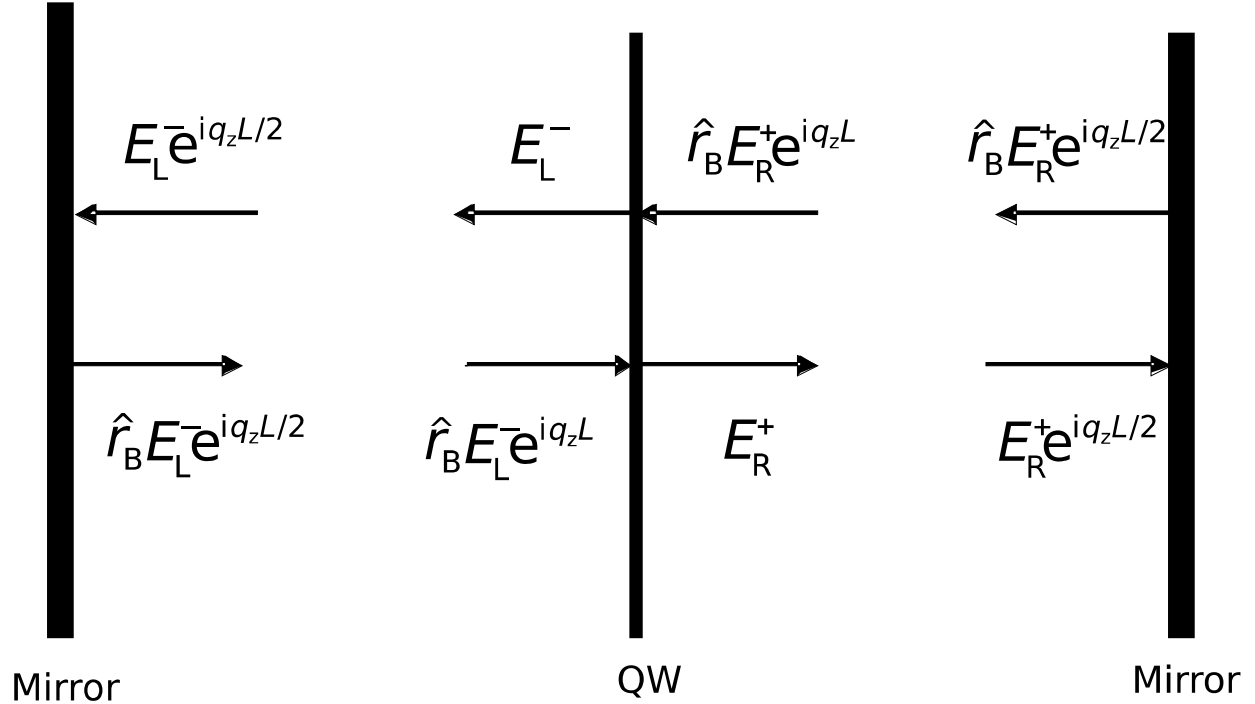}}
\caption{Schematic diagram of the microcavity with QW. Arrows show the direction of electromagnetic field propagation inside the cavity.}\label{fig:cavity}
\end{figure}

Let us now consider a QW embedded in the center of a planar
microcavity in the external magnetic field parallel to the
structure growth axis, Fig.~\ref{fig:cavity}. We shall calculate
the dispersion of the exciton-polaritons in such a system using
the generalized transfer matrix formalism. Then we compare this
exact approach with a simple model of coupled oscillators where
exciton and photon states are represented by the effective
classical oscillators.

The eigenstates of the cavity can be found from the condition that
a non-zero field exists inside the cavity without any external field. We
choose the QW center as an origin $z=0$ and present the electric
field inside the cavity (but outside the QW) in the form
\begin{equation}  \label{fields}
{\rm e}^{{\rm i} k x} \left( {\bm E}^+_L {\rm e}^{{\rm i} q_z z} + {\bm E}^-_L {\rm e}^{- {\rm i} q_z z} \right)
\hspace{3 mm} \mbox{and}  \hspace{3 mm} {\rm e}^{{\rm i} k x} \left( {\bm E}^+_R {\rm e}^{{\rm i} q_z z} + {\bm E}^-_R {\rm e}^{- {\rm i}
q_z z} \right)\:,
\end{equation}
respectively, on the left and right-hand side from the QW. The fields ${\bm E}^{\pm}_{L, R}$ are interrelated by the
transfer matrix through the QW as follows
\begin{equation}  \label{gendisp}
\hat{T}_{QW} \left( \begin{array}{c} {\bm E}_L^+ \\
{\bm E}_L^- \end{array} \right) =
\left( \begin{array}{c} {\bm E}^+_R \\
{\bm E}_R^- \end{array} \right)\:.
\end{equation}
Here $\hat{T}_{QW}$ is a $4 \times 4$ matrix, its explicit form is
given later, and ${\bm E}^{\pm}_{L,R}$ are two-component columns
with components $E_{L,x}^+, E_{L,y}^+$ etc. According to Fig.~3 ${\bm E}_L^+$ and ${\bm
E}_R^-$  can be expressed via ${\bm
E}_L^-$ and ${\bm E}_R^+$ by
\begin{equation}  \label{inter+-}
{\bm E}_L^+ = \hat{r}_B {\bm E}_L^- {\rm e}^{{\rm i} q_z \ell}\:,\:{\bm E}_R^- =
\hat{r}_B {\bm E}_R^+ {\rm e}^{{\rm i} q_z \ell}\:,
\end{equation}
where $\ell$ is the width of the active layer and $\hat{r}_B$ is a $2 \times 2$
diagonal matrix of reflection from the Bragg mirror.
Note that, at oblique incidence ($q_x = k \neq 0$), its
components $r_{B, xx}$ and $r_{B, yy}$ are different. Equations (\ref{inter+-}) can be
interpreted as follows. A QW exciton creates the
outgoing electromagnetic waves $\bm E_L^-$ and $\bm E_R^+$ propagating to the left and to the
right, respectively. The waves reaching the Bragg mirror acquire the phase
$q_z \ell/2$, the field reflected from the mirror returns to the QW
with the phase $q_z \ell$ and the additional factors $r_{B, ii}$. Whence, we obtain equation
(\ref{inter+-}) for the incoming waves.
From Eqs.~(\ref{gendisp}), (\ref{inter+-}) we arrive at the matrix dispersion
equation
\begin{equation}
\label{eigenmodes}
\hat{T}_{QW} \left( \begin{array}{c}
\hat{r}_B {\bm E}_L^- {\rm e}^{{\rm i} q_z \ell}\\
{\bm E}_L^- \end{array} \right) =
\left( \begin{array}{c} {\bm E}_R^+ \\
\hat{r}_B {\bm E}_R^+ {\rm e}^{{\rm i} q_z \ell}
\end{array}
\right)\:.
\end{equation}

The considered structure is symmetric, therefore the solutions of Eq.~\eqref{eigenmodes} can be classified
as \emph{even} [${\bm E}_L^{\pm} = {\bm E}^{\pm}_R$ in Eq.~\eqref{eigenmodes}] and \emph{odd}
(${\bm E}_L^{\pm} = - {\bm E}^{\pm}_R$) with respect to the mirror reflection in the QW plane.
The electric field in odd solutions is uncoupled with the even lowest-exciton state,
thus these modes are purely photonic. On the contrary, the even-parity cavity mode is coupled with the QW
exciton to form mixed exciton-polariton modes. The matrix equation (\ref{eigenmodes}) is equivalent to four scalar
equations. Among them, for the waves of certain parity, only two are linearly independent. If we present the transfer
matrix in the block form
\begin{equation}
 \label{tqw:bl}
\hat T_{QW} = \left(\begin{array}{cc}
 \hat T^{(11)} & \hat T^{(12)} \\  \hat T^{(21)} & \hat T^{(22)}
\end{array} \right)\:,
\end{equation}
where $\hat{T}^{(ij)}$ are $2 \times 2$ matrices, then, for the even modes, the
dispersion equation relating the frequency $\omega$ with the in-plane wave vector $k$ can be reduced to
\begin{equation}  \label{dispersion}
{\rm det} (\hat T^{(21)} \hat{r}_B {\rm e}^{{\rm i} q_z \ell} + \hat T^{(22)} -\hat r_B) =0\:.
\end{equation}

Using the definition of reflection and transmission coefficients, Eq. \eqref{eq24}, one can
present the blocks $\hat{T}^{(ij)}$ in the form
\begin{eqnarray}
\label{eq33}
T^{(11)}_{11} &=&(t_{xy} t_{xx} t_{yx} -t_{xy} r_{xx} r_{yx} -t_{xx}^2 t_{yy} -t_{yx} r_{xx}
r_{xy} +r_{yx} t_{xx} r_{xy} +r_{xx}^2 t_{yy})/\Delta\: , \\
T^{(11)}_{12} &=& (t_{yx} t_{xy}^2 -t_{xy} r_{yy} r_{xx} -t_{xy} t_{xx} t_{yy} -t_{yx}
r_{xy}^2 +r_{yy} t_{xx} r_{xy} +r_{xy} t_{yy} r_{xx})/\Delta\:, \nonumber \\
T^{(11)}_{21} &=& (-t_{xx} t_{yy} t_{yx} +t_{xx} r_{yx} r_{yy} +t_{xy} t_{yx}^2 +t_{yy}
r_{xx} r_{yx} -r_{yx}^2 t_{xy} -r_{xx} t_{yx} r_{yy})/\Delta\:, \nonumber \\
T^{(11)}_{22} &=& (-t_{xx} t_{yy}^2 +t_{xx} r_{yy}^2 +t_{xy} t_{yx} t_{yy} +t_{yy} r_{xy}
r_{yx} -r_{yy} t_{xy} r_{yx} -r_{xy} t_{yx} r_{yy})/\Delta \:, \nonumber
\end{eqnarray}
\begin{eqnarray}
&&\hat T^{(12)} = \frac{1}{\Delta} \left(
\begin{array}{cc}
 {r_{xy} t_{yx} -r_{xx} t_{yy} } & {t_{xy} r_{xx} -t_{xx} r_{xy} } \\
{t_{yx} r_{yy} -t_{xx} r_{yx} } & {r_{yx} t_{xy} -r_{yy} t_{xx} }
\end{array}
\right)\:, \\
&&\hat T^{(21)} = \frac{1}{\Delta} \left(
\begin{array}{cc}
{r_{xx} t_{yy} -t_{xy} r_{yx} } & {r_{xy} t_{yy} -t_{xy} r_{yy} } \\
{r_{yx} t_{xx} -t_{yx} r_{xx} } & {t_{xx} r_{yy} -r_{xy} t_{yx} }
\end{array}
\right) \nonumber
\end{eqnarray}
and
\begin{equation}
 \label{t22}
\hat T^{(22)} = \frac{1}{\Delta} \left(
\begin{array}{cc}
 -t_{yy} & t_{xy}\\
  t_{yx} & -t_{xx}
\end{array}
\right) \:,
\end{equation}
where $\Delta = t_{xy} t_{yx} -t_{xx} t_{yy}$.

In general, Eq.~(\ref{dispersion}) has four solutions for each
incidence angle which corresponds to four exciton-polariton
dispersion branches. The polarization of polariton eigenmodes is
determined by the electric-field complex amplitudes $E_{x}$ and $E_{y}$. The degree of
polariton circular polarization is given by the standard expression~\cite{LandauLif}

\begin{equation} \label{eq41}
\rho_c =  \frac{ 2 {\rm Im} \{ E^{\ast}_x E_y \}}{  |E_x|^2 + |E_y|^2 }\:.
\end{equation}

In the absence of an external magnetic field the cross-polarized reflection and transmission coefficients
vanish, $r_{xy}=r_{yx}=t_{xy}=t_{yx}=0$, and Eq.~(\ref{dispersion}) reduces to a couple of
independent equations for TE- and TM-polarized polariton
modes satisfying the dispersion equation~\cite{ivchbook}:
\begin{equation}
\label{eq42}
(2r_{ii}+1) r_{B,ii} {\rm e}^{{\rm i} q_z \ell} = 1\:,
\end{equation}
where $i=x$ (TM) or $i=y$ (TE). In this case exciton-polariton modes have definite linear
polarizations.

In order to demonstrate underlying physics of the light-matter coupling we also put forward a
simplified model of four coupled oscillators representing the TE- and TM- polarized optical modes of empty cavity and two
excitonic modes split by the Zeeman interaction. The effective
dispersion equation can be written as
\begin{equation}
\label{eq44}
\det \left( {{\begin{array}{*{20}c}
 {\hbar \omega_0(k) - E} \hfill & {{\rm i} g \mu _B B/2} \hfill & {V /2}
\hfill & 0 \hfill \\
 {- {\rm i} g \mu _B B/2} \hfill & {\hbar \omega_0(k) - E} \hfill & 0
\hfill & {V /2} \hfill \\
 {V /2} \hfill & 0 \hfill & {E_{ph}^{\left( {\rm TE} \right)} (k)-E} \hfill &
0 \hfill \\
 0 \hfill & {V /2} \hfill & 0 \hfill & {E_{ph}^{\left( {\rm TM} \right)}
(k)-E} \hfill \\
\end{array} }} \right)=0 \:.
\end{equation}
Here $V$ is the Rabi splitting determined by the system
parameters, $E_{ph}^{\left( {\rm TE} \right)} (k)$, $E_{ph}^{\left(
{\rm TM} \right)} (k)$ are dispersions of the TE and TM bare-cavity
photons. We neglected the longitudinal-transverse splitting of the exciton.

The difference between this approach and the rigorous formalism
presented above consists in (i) neglecting of the coupling of four
polariton modes with all other light modes in the cavity and (ii)
disregarding of the polarization dependence of the light-matter
coupling constant $V_R$~\cite{goupalov}.

\begin{figure}[htbp]
\centerline{\includegraphics[width=\linewidth]{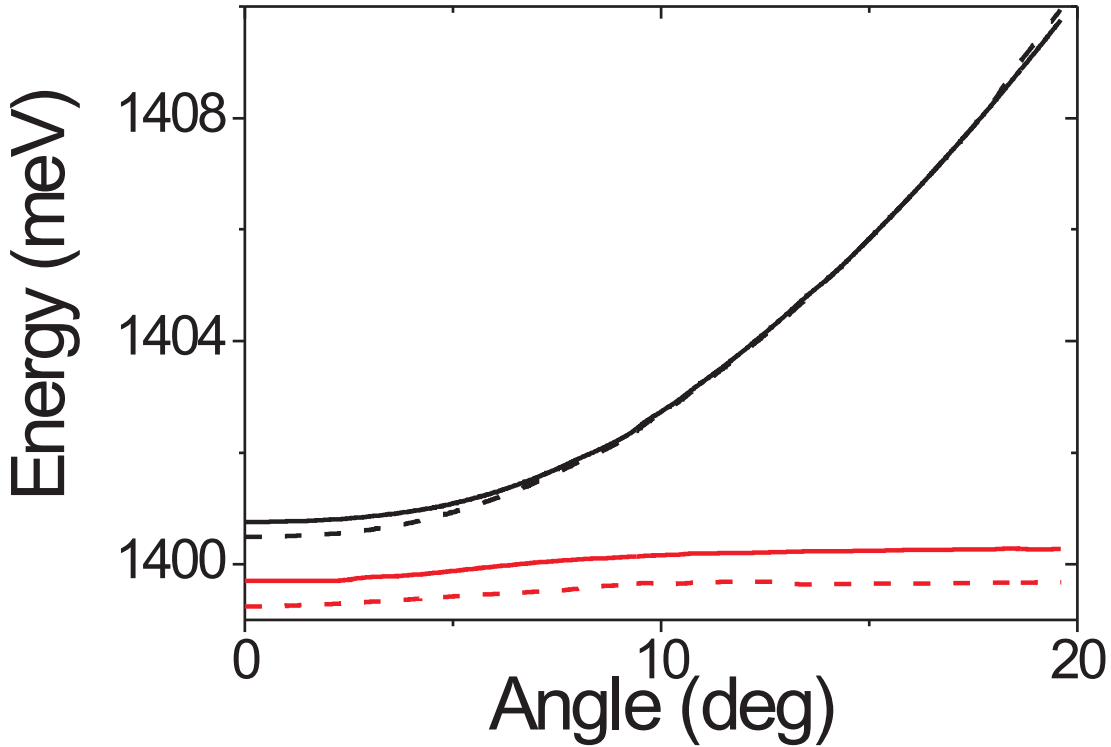}}
\caption{Dispersion of exciton-polaritons in the cavity in strong magnetic
field $B=11.5$~T ($g\mu_B B = 0.6$~meV) calculated for the exact Eq.~\eqref{dispersion}. Red and black mark lower and upper polariton branches, correspondingly, and solid and dashed lines correspond to different polarizations.
Cavity width is $\lambda=246$ nm. Other parameters the same as in Fig. 1.}\label{fig:dispersion}
\end{figure}

Figure~\ref{fig:dispersion} shows the exact dispersion of cavity
polaritons for an artificially high value of the magnetic field
11.5 T allowing to make visible all peculiarities of the
dispersion. This field corresponds to the Zeeman splitting of
0.6~meV.  We consider a $\lambda$-cavity with 10 nm GaAs QW
(parameters are the same as in Fig.~\ref{fig:faraday}). The
distributed Bragg reflectors (DBRs) are typical 20-period
GaAs/Al$_{0.18}$Ga$_{0.82}$As mirrors. Approximate dispersion
curves calculated from Eq. \eqref{eq44} are not shown, because
they do not differ from the exact one in the range of angles
considered. Four branches of exciton-polaritons having a
characteristic non-parabolic dispersion are clearly seen. The
anticrossings take place between branches having the same circular
polarization. In such a strong field, the polariton eigenstates
are almost completely circularly polarized, so we do not show
their polarization degree in this case.

\begin{figure}[htbp]
\centerline{\includegraphics[width=\linewidth]{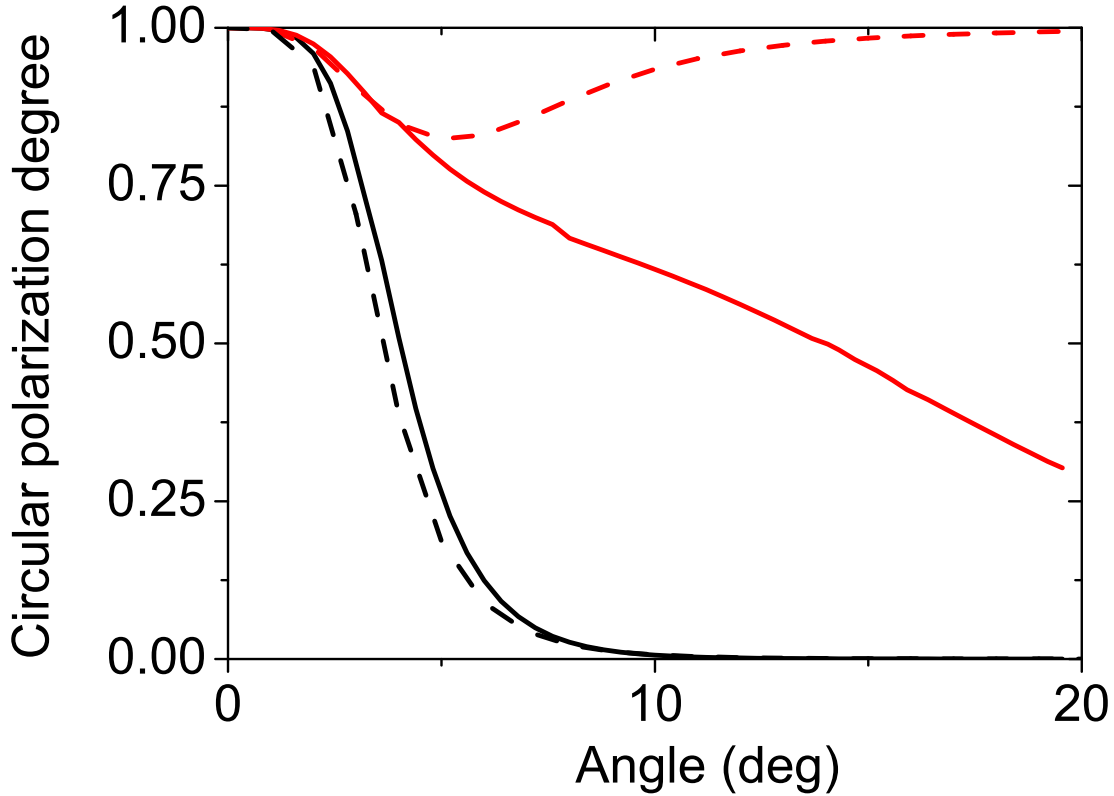}}
\caption{Circular polarization degree in the case of magnetic field $B=0.04$~T ($g\mu_B B = 2\times 10^{-3}$~meV)
(solid line -- exact model, dashed line -- approximate model; red color shows exciton-like
branch and black color shows photon-like branch). Only branches with positive polarization
degree are shown.
}\label{fig:dispersionpolarization}
\end{figure}

However, we do show the the circular polarization degree of the
cavity eigenstates in the exact and approximate models for a
small magnetic field of 0.04 T (Zeeman splitting is 2~$\mu$eV) in
figure~\ref{fig:dispersionpolarization}. The results match well
except for the exciton-like branch (red), the difference coming
from the fact that in the exact model the exciton eigenstates are
circularly polarized, whereas in the approximate model they are
linearly polarized (their TE-TM splitting is taken into account). 
The circular polarization degree has been
calculated using Eq.~\eqref{eq41} in both models, the difference
being that in one case the complex amplitudes are the eigenvectors
of exact system Eq.~\eqref{dispersion} and in the other case they are
the eigenvectors of approximate system Eq.~\eqref{eq44}. For such
a small field, the splittings that appear in the polaritonic
branches, are invisible relative to the Rabi splitting, therefore
we do not show the corresponding dispersions.

\section{Kerr and Faraday rotation in quantum microcavities}

In this section we consider Kerr and Faraday effects in quantum microcavities. It is assumed that $s$ or $p$ polarized light is incident on a microcavity from the vacuum and the rotation of polarization plane of reflected (Kerr effect) and transmitted (Faraday effect) wave is monitored as a function of incidence angle.

The Kerr and Faraday rotation angles can be determined once the matrices of amplitude reflection (transmission) coefficients $r_{c,ij}$ ($t_{c,ij}$) of the whole microcavity are known. The latter can be found from the microcavity transfer matrix $\hat T_{\rm qmc}$ which, in accordance with its definition, can be recast as a product
\begin{equation}
 \hat T_{\rm qmc} = \hat T_{B} \hat T_l \hat T_{QW} \hat T_l \hat T_{B},
\end{equation}
where $\hat T_B$ is the transfer matrix through the Bragg mirror, $\hat T_l$ is the transfer matrix through a half of the cavity and $T_{QW}$ is the quantum well transfer matrix, determined in Eqs. \eqref{tqw:bl}, \eqref{eq33}--\eqref{t22}. The transfer matrix through the homogeneous layer of the thickness $l$ is diagonal and given by
\begin{equation}
 \hat T_l = \left(
\begin{array}{cc}
 e^{\mathrm i q_z \ell/2} \hat I & 0 \\
0 & e^{-\mathrm i q_z \ell/2} \hat I
\end{array}
\right),
\end{equation}
where $\hat I$ is the $2\times 2$ unit matrix.

\begin{figure}[htbp]
\centerline{\includegraphics[width=\linewidth]{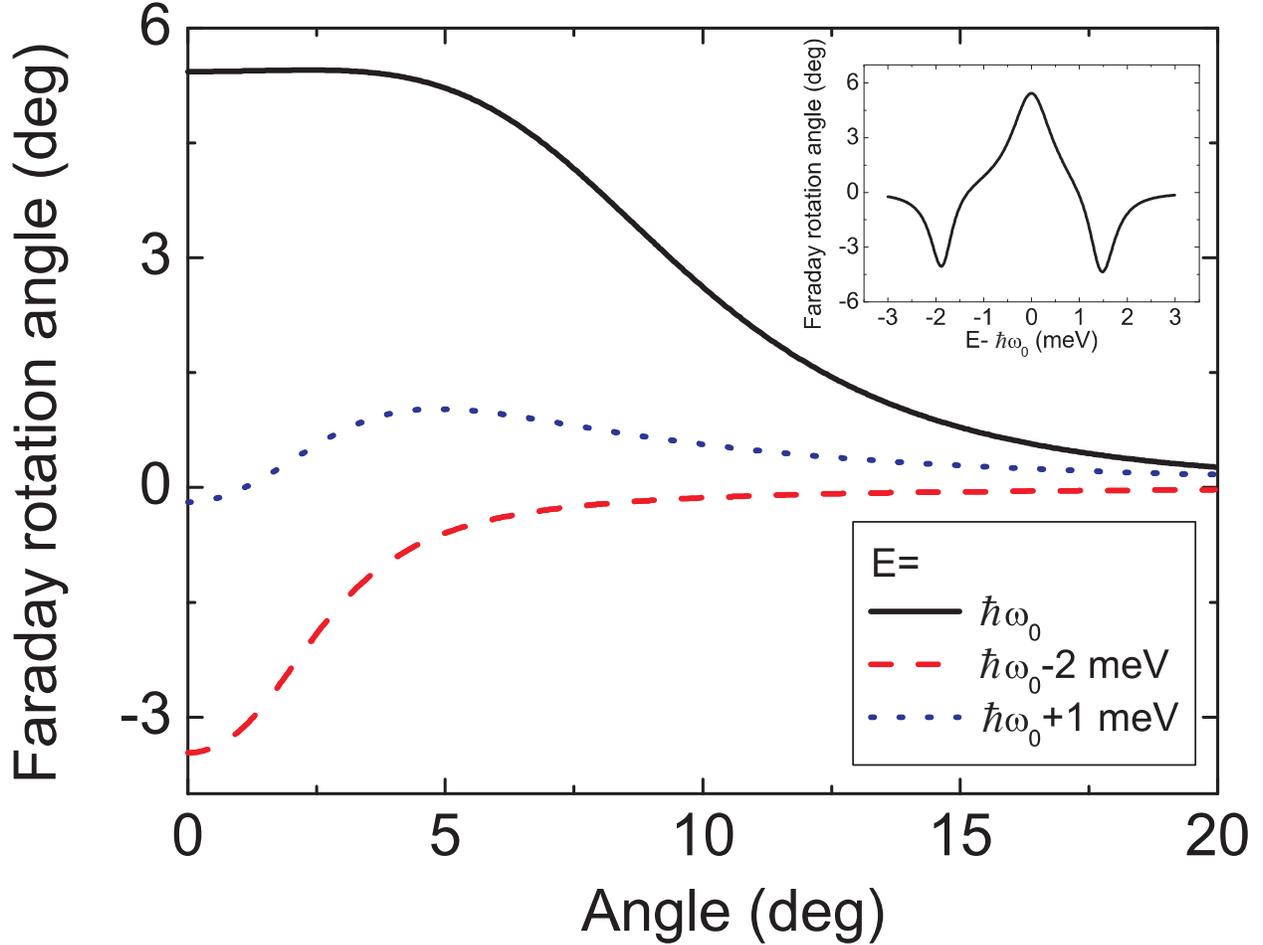}}
\caption{Faraday rotation angle as a function of the incidence angle for the $s$ polarized light. The QW and cavity parameters are the same as before. The Zeeman splitting is 0.1~meV. Inset shows energy dependence of Faraday rotation angle at the same parameters at normal light incidence.
}\label{fig:cavityfaraday}
\end{figure}

For example, the transmission and reflection coefficients in the $s$ polarization can be determined from the following system of equations
\begin{equation}
 \label{spol}
\hat T_{qmc} 
\left(
\begin{array}{c}
 0 \\
1\\
r_{c,xy}\\
r_{c,yy}
\end{array}
\right) = 
\left(
\begin{array}{c}
 t_{c,xy} \\
t_{c,yy}\\
0\\
0
\end{array}
\right).
\end{equation}

Figure~\ref{fig:cavityfaraday} shows calculated Faraday rotation angle at oblique incidence as a function of incidence angle. An inset in the figure presents the Faraday rotation angle as a function of the energy. The function shows a non-monotonous behaviour reflecting the resonant character of light reflection and transmission in microcavities. The figure clearly demonstrates that Faraday rotation angles are almost by an order of magnitude larger than those for the Faraday rotation in the case of a single quantum well. This enhancement can be interpreted as a result of multiple passage of a photon inside a cavity~\cite{Kavokin:1997}. 

Namely, consider a cavity mode as a beam of light travelling back and forth inside the cavity. Let $N$ be an average number of photon round trips between the Bragg mirrors made during its life-time ($N\sim Q$, where $Q$ is the cavity quality factor). At each trip the polarization plane of the photon exhibits a rotation by $\delta\theta$ and the Faraday effect is strongly enhanced. The observable Faraday rotation angle is smaller than $N\delta\theta$ because there is a finite probability for the photon to escape cavity after any number of passages~\cite{Kavokin:1997}. The absolute value of the Faraday rotation angle decreases with the increase of the incidence angle, due to the decrease of the polariton lifetime in the cavity. In average, the polaritons make less rount-trips inside the cavity at oblique angles than at normal angle.
We note that the Kerr rotation angles (observed in reflection geometry) are smaller than the Faraday rotation angles because the light reflection from the microcavity is dominated by a surface reflection of the Bragg mirror.

\section{Conclusions}

In conclusion, we have analyzed the dispersion of the exciton-polaritons in a microcavity subjected into the external magnetic
field taking in account both Zeeman splitting of the exciton and
TE-TM splitting of the photonic modes. We have shown that the
polarization of the polariton eigenstates is neither linear nor
circular, but elliptical, in general. It is very sensitive to the
polariton in-plane wave vector as well as to the detuning between
the exciton and photon modes. 

We have also studied theoretically the Faraday rotation in the planar microcavities. We found that the giant Faraday rotation due to the multiple passages of light across the quantum well takes place at normal incidence, while this effect is reduced at oblique incidence angles. 

\acknowledgements{We acknowledge support from the EU projects
ANR Chair of Excellence and STIMSCAT FP6-517769 and the Royal Society Joint
International project. M.M.G. and E.L.I. were partially supported
by RFBR, Programmes of RAS and Dynasty Foundation-ICFPM.}

\end{document}